\pdfoutput=1
\documentclass{JINST}

\usepackage{overpic}
\usepackage{booktabs}

\title{Simulation and performance of an artificial retina for $\mathbf{40}$~MHz track 
reconstruction}

\author{A.~Abba$^b$, F.~Bedeschi$^c$, M.~Citterio$^b$, F.~Caponio$^b$,
A.~Cusimano$^b$, A.~Geraci$^b$,   
P.~Marino$^d$\thanks{Corresponding author.},
M.~J.~Morello$^d$, N.~Neri$^b$, G.~Punzi$^c$,
A.~Piucci$^c$, L.~Ristori$^{ce}$, F.~Spinella$^c$,
S.~Stracka$^d$ and D.~Tonelli$^a$. \\
\llap{$^a$}CERN\\
  385 Route de Meyrin, Geneva, Switzerland\\
\llap{$^b$}Politecnico and INFN-Milano,\\
 Via Celoria 16, 20133, Milano, Italy\\
 \llap{$^c$}University and INFN-Pisa,\\
 L.go Bruno Pontecorvo 3, 56127, Pisa, Italy\\
  \llap{$^d$}Scuola Normale Superiore and INFN-Pisa,\\
Piazza dei Cavalieri 7, 56126, Pisa, Italy\\
  \llap{$^e$}Fermilab,\\
Wilson and Kirk Rd, Batavia, IL 60510, USA\\
E-mail: \email{pietro.marino@pi.infn.it}}

\abstract{We present the results of a detailed simulation of the artificial 
retina pattern-recognition algorithm, designed to reconstruct events
with hundreds of charged-particle 
tracks in pixel and silicon detectors at LHCb with LHC crossing frequency
of $40\,\rm MHz$. 
Performances of the artificial retina algorithm are assessed using
the official Monte Carlo samples of the LHCb experiment. 
We found performances for the retina pattern-recognition algorithm
comparable with the full LHCb reconstruction algorithm.}

\keywords{Pattern recognition; Trigger algorithms}

\begin{document}


\section{Introduction}\label{sec:intro}

Higher LHC energy and luminosity increase the challenge 
of data acquisition and event reconstruction in the LHC experiments.
The large number of interaction for bunch crossing (pile-up) greatly reduce
the discriminating power of usual signatures, 
such as the high transverse momentum leptons or the high transverse missing energy.
Therefore real-time track reconstruction is needed to quickly select
potentially interesting events for higher level of processing.
Performing such a task
at the LHC crossing rate is a major challenge
because of the large combinatorial and the size of the associated information flow
and requires unprecedented massively parallel pattern-recognition algorithms.
For this purpose we design and test a  
neurobiology-inspired pattern-recognition algorithm 
well suited for such a scope: the \emph{artificial retina algorithm}.

\section{An \emph{artificial retina} algorithm }\label{sec:retina}

The original idea of an \emph{artificial retina} tracking algorithm was inspired
by the mechanism of visual receptive fields in the mammals eye~\cite{retinaNIM}.
Experimental studies have shown neurons
tuned to recognize a specific shape on specific region of the retina
(``receptive field'')~\cite{hubel1962receptive, priebe2012mechanisms}.
The strength of the response of each neuron to a stimulus is proportional to how 
close 
the shape of the stimulus is to the shape for which the neuron is tuned to. 
All neurons react to a stimulus, each with different strength, 
and the brain obtains a precise information of the received stimulus 
performing some sort of interpolation between the responses of neurons.

\begin{figure}[tbp] 
\centering
\includegraphics[width=.8\textwidth]{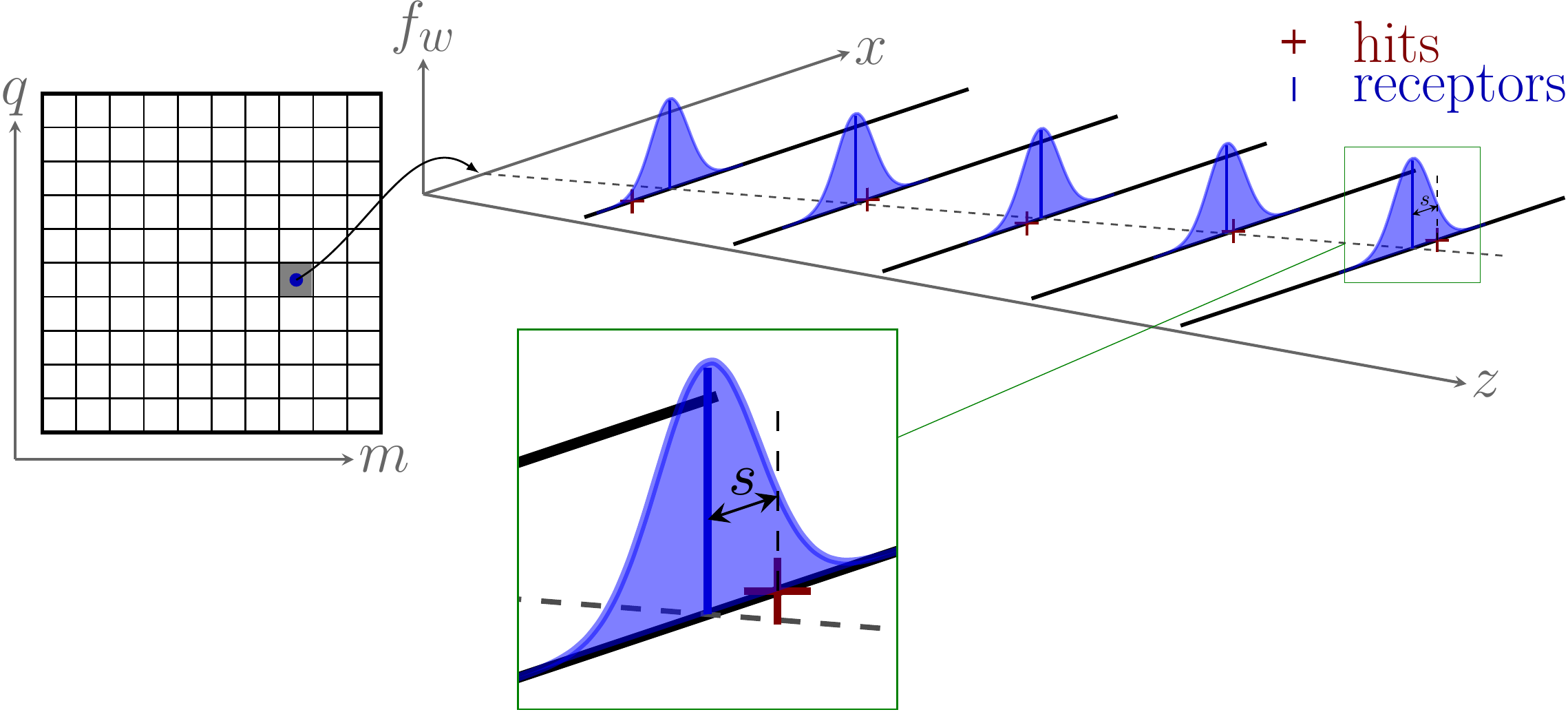}
\caption{Schematic representation of the detector mapping (more details in the text). 
The grid in parameter space (left) and
the corresponding  receptors in the detector (right).}
\label{fig:mapping2D}
\end{figure}

The retina concepts can be geared toward track reconstruction.
Assuming a tracking detector
made by a set of parallel layers, providing the measurement of a single spatial coordinate $(x)$, 
and a detector volume without any magnetic field.
Thus trajectories of charged particles
are straight lines, intersecting detector layers, and are identified 
by two parameters, \emph{e.~g.} $(m, q)$, where $m$ is the angular coefficient and $q$ is 
the intersection with the $x$-axis in the $(z,x)$-plane.
 We discretize the space of track parameters, $(m,q)$, into \emph{cells}, 
representing the receptive fields of the  visual system.
The centre of each cell identifies a track in the detector space, that intersects
detector layers in spatial points that we call \emph{receptors}. 
Therefore each $(m_i,q_j)$-cell of the parameter space 
corresponds to a set of receptors $\{x_k^{ij}\}$,
where $k=1, \ldots, n$ runs over the detector layers, as shown in figure~\ref{fig:mapping2D}. 
This procedure is called \emph{detector mapping} and it is done for all the cells of the track parameter space, 
covering all the detector acceptance.
For each incoming hit, the algorithm computes
the excitation intensity, \emph{i.~e.} the response of the receptive field,
of each $(m_i, q_j)$-cell as follows:
\begin{equation}
\label{eq:responsecell}
R_{ij} = \sum_{k,\,r} \exp\Big(-\frac{s^2_{ijkr}}{2\sigma^2}\Big),
\end{equation}
using the distance 
\begin{equation}
\label{eq:distancecell}
s_{ijkr} = \overline{x}_{k,r} -x_k^{ij},
\end{equation}
where $\overline{x}_{k,r}$ is the $r$-th hits on the detector layer $k$, while 
$\sigma$ is a parameter of the retina algorithm, that it can be adjusted 
to optimize the sharpness of the response of the receptors.

After all hits are processed, tracks are identified as local maxima 
over a threshold in the space of track parameters.
\begin{figure}[tbp]
\centering
\includegraphics[width=.48\textwidth]{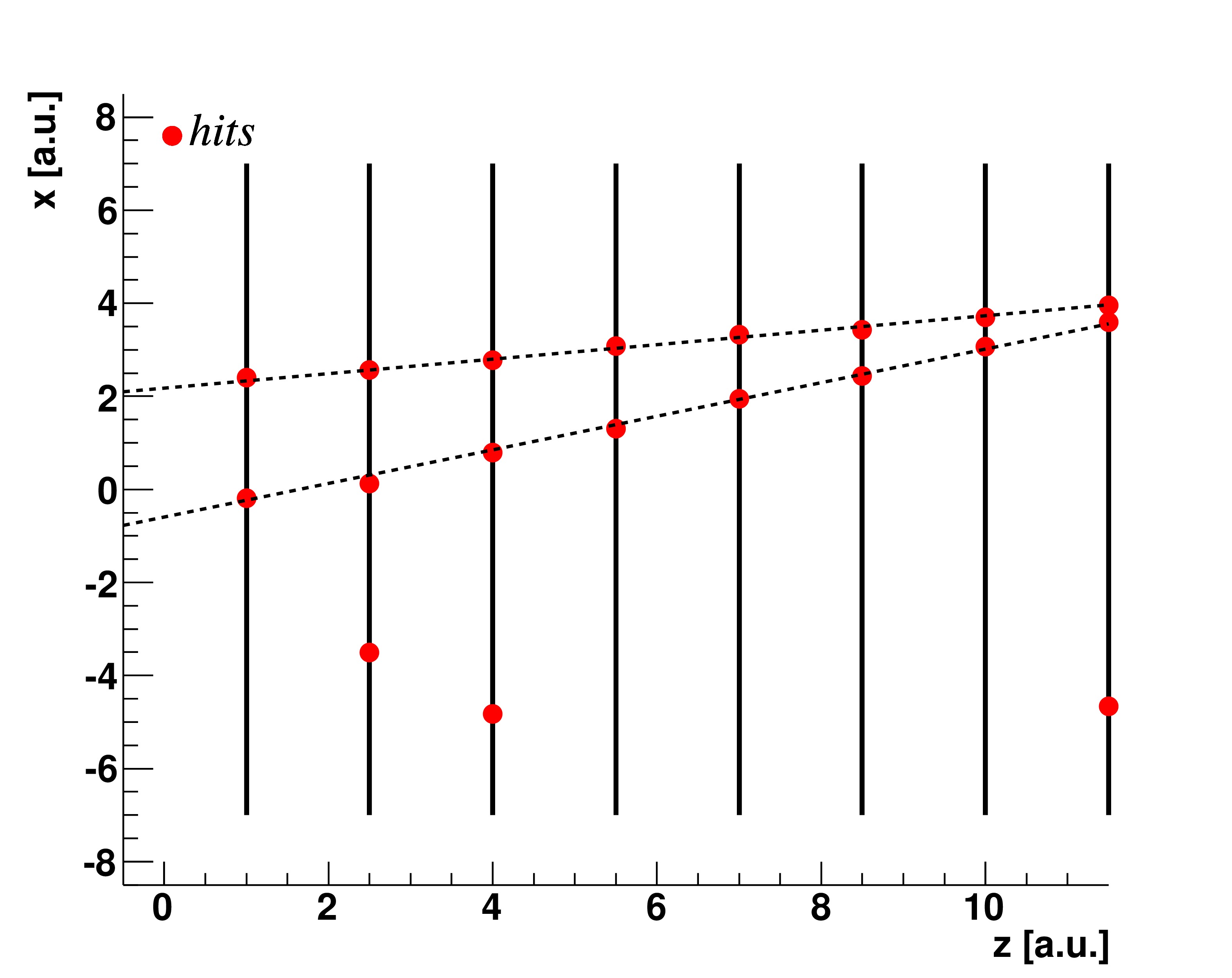} \\
\includegraphics[width=.48\textwidth]{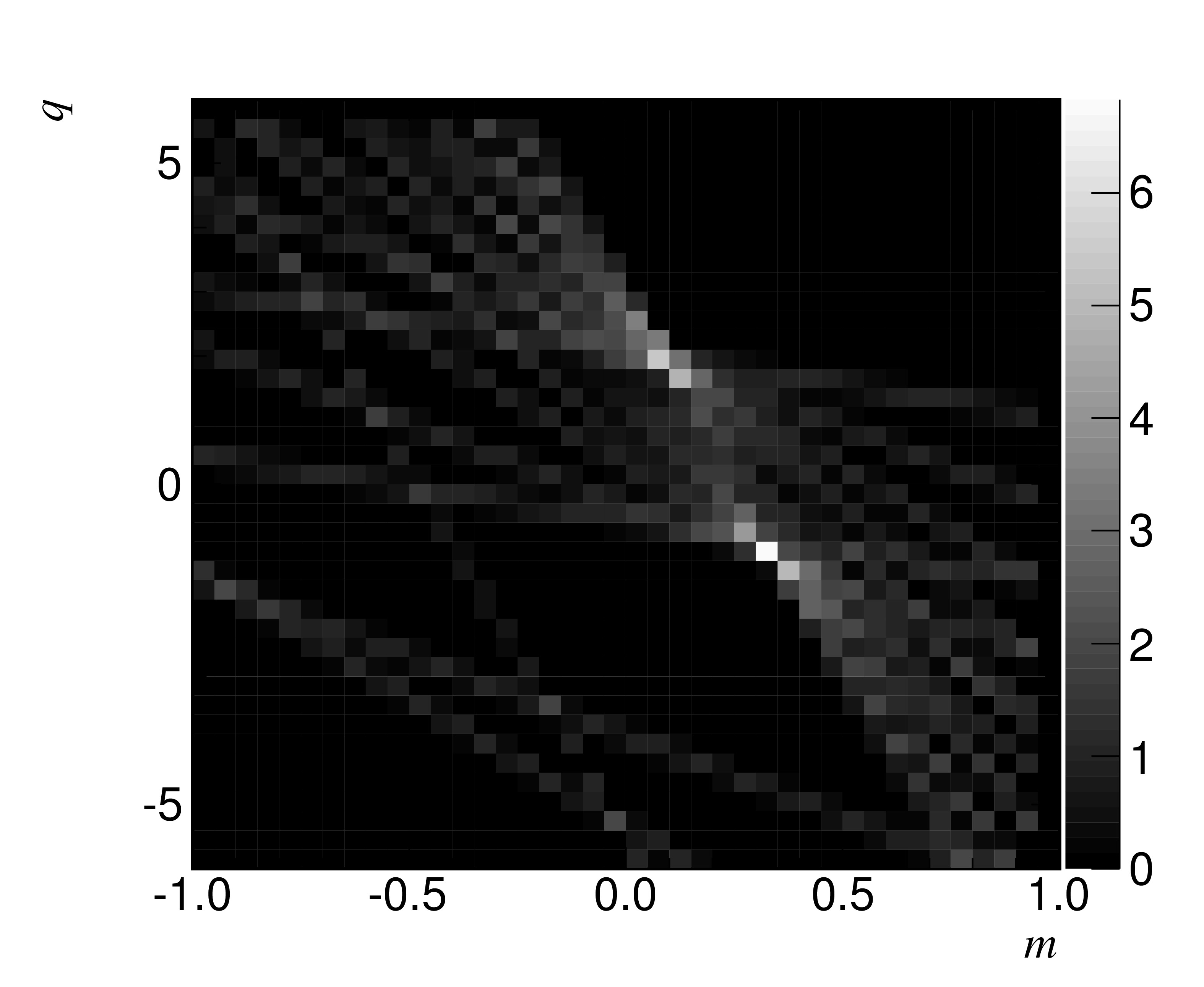}
\includegraphics[width=.48\textwidth]{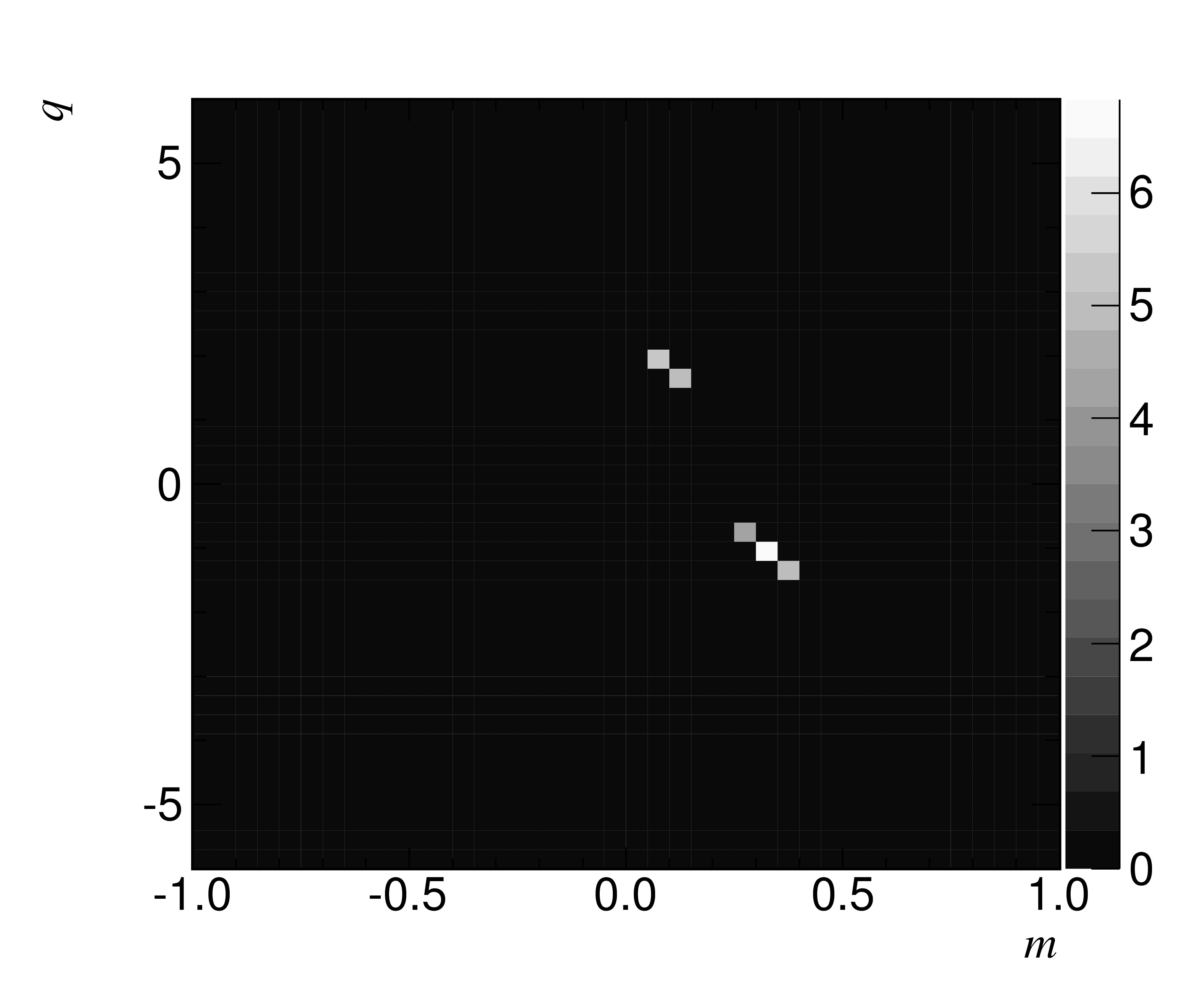}
\caption{Response of the retina (bottom left) to a particular event (top).
Bottom right, tracks reconstructed by the retina (over-a-threshold maxima).}
\label{fig:response_retina}
\end{figure}
Averaging over nearby cells it is possible to extract track parameters 
with a significant better resolution
than the available cell granularity. 
Hence, track parameters are extracted
performing the centre of mass of the cluster.

\section{Retina algorithm in a real HEP experiment}

In a real HEP detector, 
the geometry and the topology of the events are quite different from 
the simple case described in section~\ref{sec:retina}.
For example, trajectories of charged particles are not straights lines, 
because of the presence of the magnetic field (necessary to measure their momenta),
and are affected by multiple scattering and detector noise effects.
In addition, the retina response to realistic high track
multiplicity LHC events should be studied, to understand if the size of
the retina in terms of number of cells, at fixed desired tracking
performances, is limited, in order to check if a realistic implementation
using current available technology is feasible or not. 
Therefore an accurate study of the retina algorithm in a realistic environment is necessary. 

\begin{figure}[tbp]
\centering
\includegraphics[width=.6\textwidth]{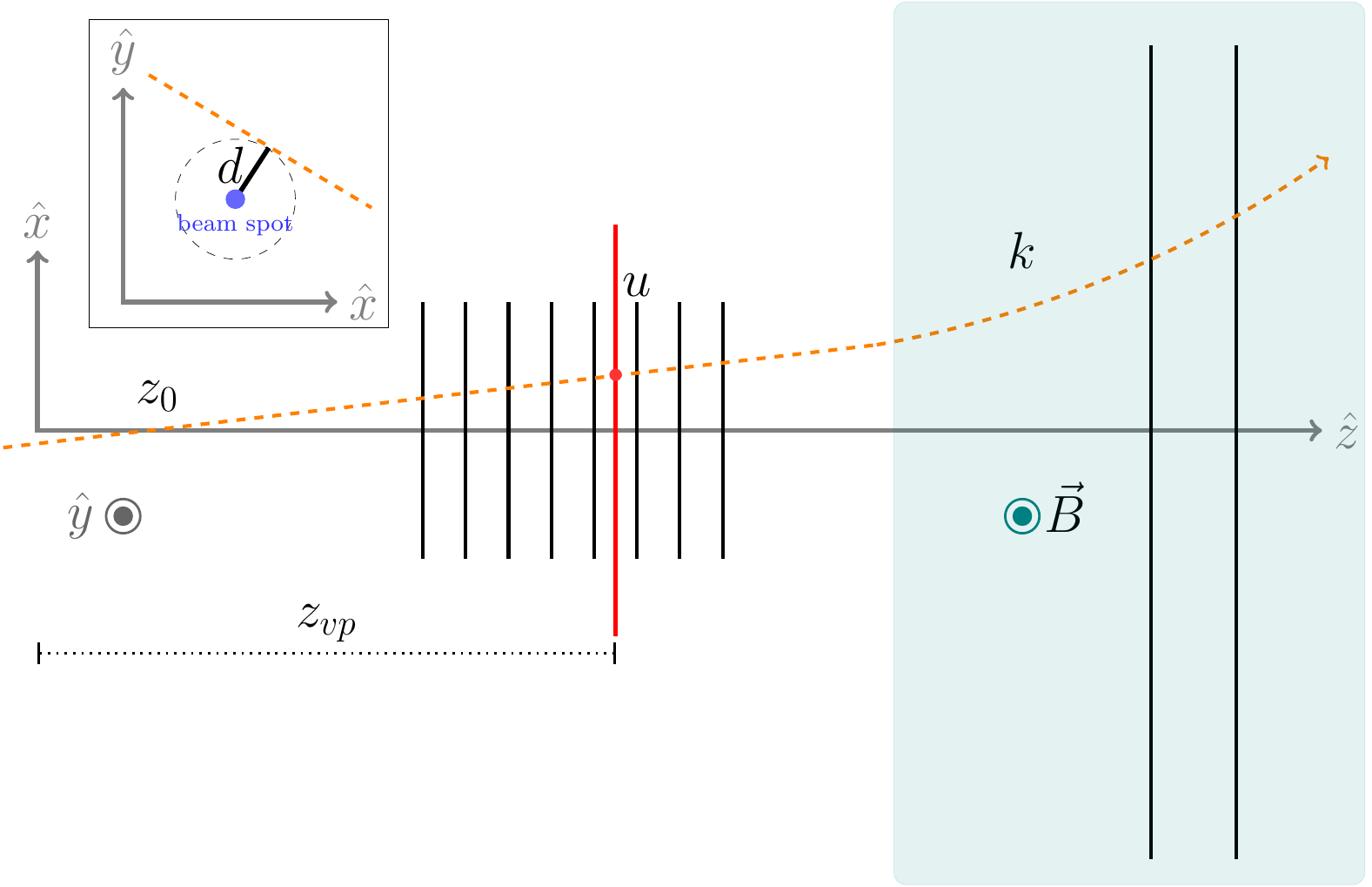}
\caption{Schematic representation of the tracker geometry (not drawn to scale). 
There is not magnetic field in the VELO detector, while the UT detector is sink into the weak fringe magnetic field
of the upstream $4\,\mathrm{Tm}$ magnet~\cite{LHCb-now}.}
\label{fig:detector_scheme}
\end{figure}

\begin{figure}[tbp]
\centering
\includegraphics[width=.6\textwidth]{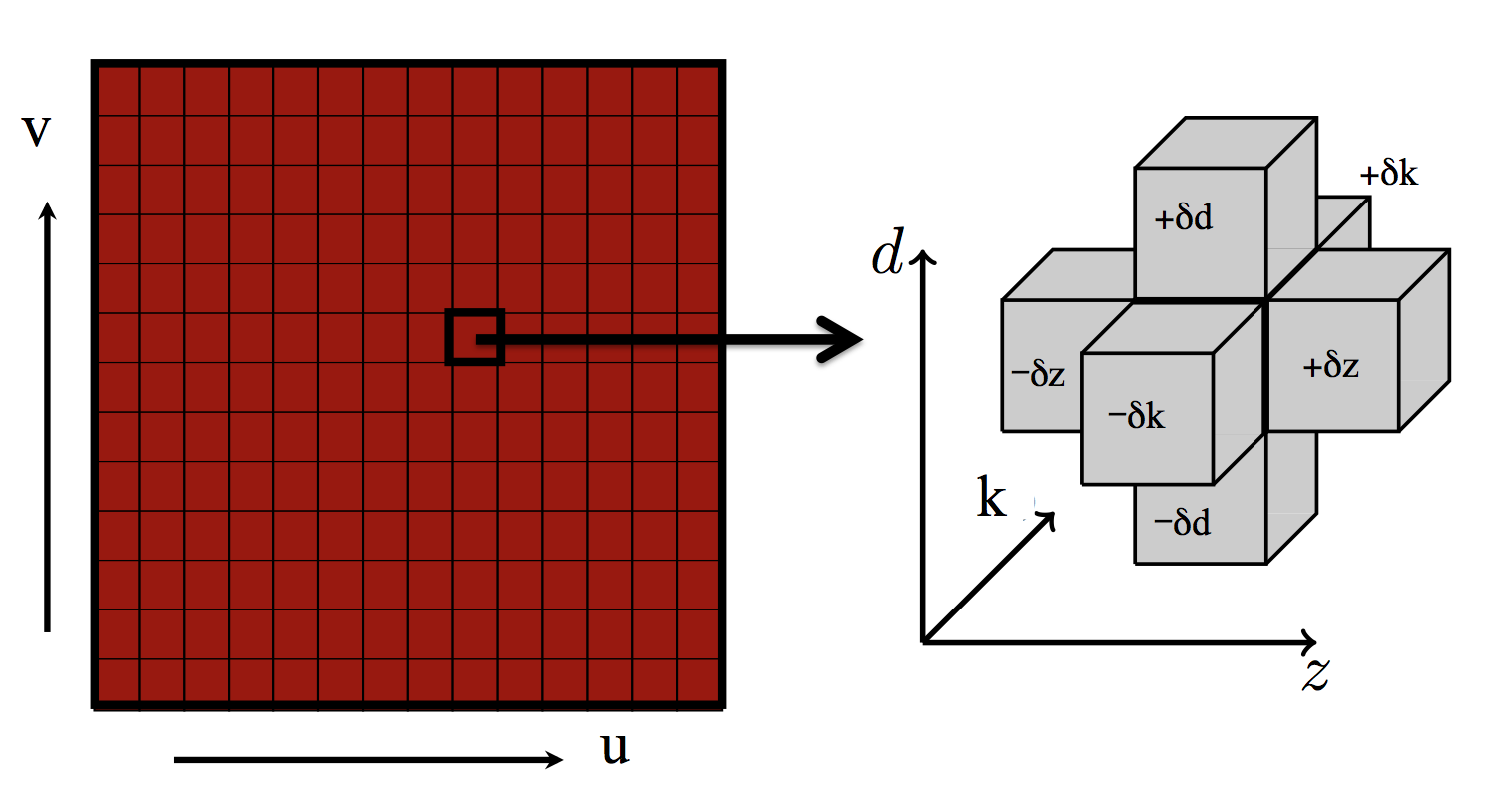}
\caption{Schematic representation of the lateral cells implementation.}
\label{fig:subcells}
\end{figure}

We chose to focus our work on the LHCb-Upgrade detector, 
where tracking plays a particularly important role
in collecting enriched samples of flavored events.
LHCb is a single-arm spectrometer covering the pseudorapidity range 
$2<\eta<5$, specialized to study heavy flavored events.
LHCb-Upgrade~\cite{LHCb-Upgrade} is a major upgrade
of the current LHCb experiment~\cite{LHCb-now} and it will
start data taking after the Long Shutdown 2 (LS2) of the LHC, in 2020, 
at the instantaneous luminosity of $3\times 10^{33}$~cm$^{-2}$s$^{-1}$.
All the sub-detectors will be read at $40\,\rm MHz$, 
allowing a complete event reconstruction at the LHC crossing rate.
To benchmark the retina algorithm, we decided
to perform the first stage of the LHCb-Upgrade tracking sequence~\cite{VELOUT},
performing the track reconstruction using the information of 
only two sub-detectors, placed upstream of the magnet:
the VErtex LOcator (VELO), a silicon-pixel detector~\cite{VELOTDR} and
the Upstream Tracker (UT)~\cite{UTTDR}, a silicon mini-strip detector.
We used the last eight forward pixel layers of the 
VELO and the two axial layers of the UT.
A sketch of the chosen configuration is reported in figure~\ref{fig:detector_scheme}.
The 3D trajectory of a charged particle is
identified by five parameters. We arbitrarily chose:
\begin{description}
	\item[{$\mathbf{u,v}$}] spatial coordinate of the intersection point of the track with a ``virtual plane''
	perpendicular to the $z$-axis, placed to a distance $z_{vp}$ from the origin of the 
	coordinate system (red plane in figure~\ref{fig:detector_scheme});
	\item[$\mathbf{d}$] signed transverse impact parameter, the distance 
	of the closest approach to the $z$-axis;
	\item[$\mathbf{z_0}$] $z$-coordinate of the point of the closest approach to the $z$-axis;
	\item[$\mathbf{k}$] signed curvature, defined as $q/\sqrt{p_x^2 + p_z^2}$, where $q$ is the 
	charge of the particle, and $p_x$ and $p_z$ are the coordinate momentum of the track 
	perpendicular to the main magnetic field direction~$(\vec{B} = B\hat{y})$.
\end{description}
Following the full-fledged approach of the retina algorithm,
how discussed in section~\ref{sec:retina},
the five dimensional track-parameter space $(u,v,d,z_0,k)$
has to be discretize into cells.
However, a 5-dimensional parameter space may easily lead to
an impracticable number of cells.
We chose the approach of selecting only 2 dimensions 
as ``main'' parameters, to be counted on for pattern recognition of tracks, 
leaving the other parameters to be treated as ``perturbations''.
This approach is supported by the detector geometry, since 
the intensity of the magnetic field is negligible in the VELO volume,
and only  a weak fringe field is present in  the UT volume. Tracks can be approximated as straight lines and 
 identified  by only two main parameters $(u,v)$, assuming they
 originate from a single point.   
Variations of ``secondary'' ($d,z_0,k$) parameters affect only marginally the shape
of the 2-dimensional retina cluster in the ($u,v$) space  allowing to
fully perform pattern recognition 
without any  degradation in performances.  \\
The $(u,v,d,z_0,k)$ parameter space is then divided into cells, but a
fine grid is used in the $(u,v)$ plane, while only 3 bins are taken in the other directions,
named ``lateral cells''.
For each main cell  $(u,v, 0, 0, 0)$, we calculate two subcells for each secondary parameter
$(u,v,\pm \delta d, \pm \delta z_0, \pm \delta k)$, for a total of six ``lateral cells'', 
see figure~\ref{fig:subcells}.
Once a cluster over a threshold (a track) is found  ($u_m,v_n$),   
the $(u,v)$ track parameters are extracted finding the centre of mass of a $3 \times 3$-cells square:
\begin{equation}
	\overline{u} = \frac{\sum_{ij} u_i W_{ij} }{ \sum_{ij} W_{ij}}; \qquad 
	\overline{v} = \frac{\sum_{ij} v_i W_{ij} }{ \sum_{ij} W_{ij}};
\end{equation}
with $i =(m-1, m, m+1)$ and $j = (n-1, n, n+1)$. $W_{ij}$ is the 
excitation intensity of the correspondent $(u_i,v_j)$ cell.
The $(d, z_0, k)$ parameters are instead calculated by interpolating the response 
of the two lateral subcells corresponding to the considered parameter, for instance:
\begin{equation}
	\overline{d} = \frac{ \sum_{i} d_i W_{i} }{ \sum_i W_i }
\end{equation}
where $d_i$ assumes the values $(-\delta d, 0, \delta d)$, and $W_i$
is the excitation intensity of the lateral cell corresponding to the $d_i$ cell.
The same holds for $\overline{z}_0$ and $\overline{k}$.

Because of the forward detector geometry and the topology of physics events, 
 tracks are not uniformly distributed  in the space of parameters, leading to an inefficient cells distribution, 
if uniformly sampled. 
We have performed appropriate non-linear transformation
of coordinate measured on the virtual plane,
to achieve a track distribution which is uniform in the $(u,v)$ space.
It may be noted that this transformation has a close similarity 
with what the real retina achieves 
with the non-uniform distribution of photoreceptors in the fovea.
The non-linear transformation is derived using the probability distribution of tracks
in the parameter space evaluated from a minimum-bias data sample from
the official LHCb-Upgrade simulation.

\section{Retina performances}

The algorithm described in this paper is implemented in a software package called \emph{Retina Simulator}. 
The simulator is completely written in \texttt{C++}~\cite{Cpp}
and it uses the ROOT data analysis framework~\cite{ROOT}.
As well as to evaluate the performances of the algorithm, in a real
experiment, the retina simulator has been also developed to drive
and assist  the hardware implementation
of the retina algorithm on FPGAs (more details in~\cite{talk_Diego}).
All the features discussed in the previous section are implemented in the simulator. 
In addition, it 
interfaces the official LHCb simulation, being able
to process simulated LHCb events.
The receptors banks are extracted from the LHCb simulation.
For each cell of the parameter space a particle (a muon), with parameters corresponding
to the central values of the cell, is ``shot'' through the detector.
Intersections of this sample track with the detector layers are the receptors of the cell.%

The main $(u,v)$-subspace is divided into $22\,500$ cells, 
a granularity $\mathcal{O}(100)$
larger than the maximum expected number of tracks in a typical LHCb-Upgrade event.
The steps chosen for the lateral cells are $\delta d = 1\,\mathrm{mm}$, $\delta z_0 = 150\,\mathrm{mm}$ and 
$\delta k = 1\,(\mathrm{GeV}/c)^{-1}$.
\begin{figure}[tbp]
\centering
\includegraphics[width=.53\textwidth]{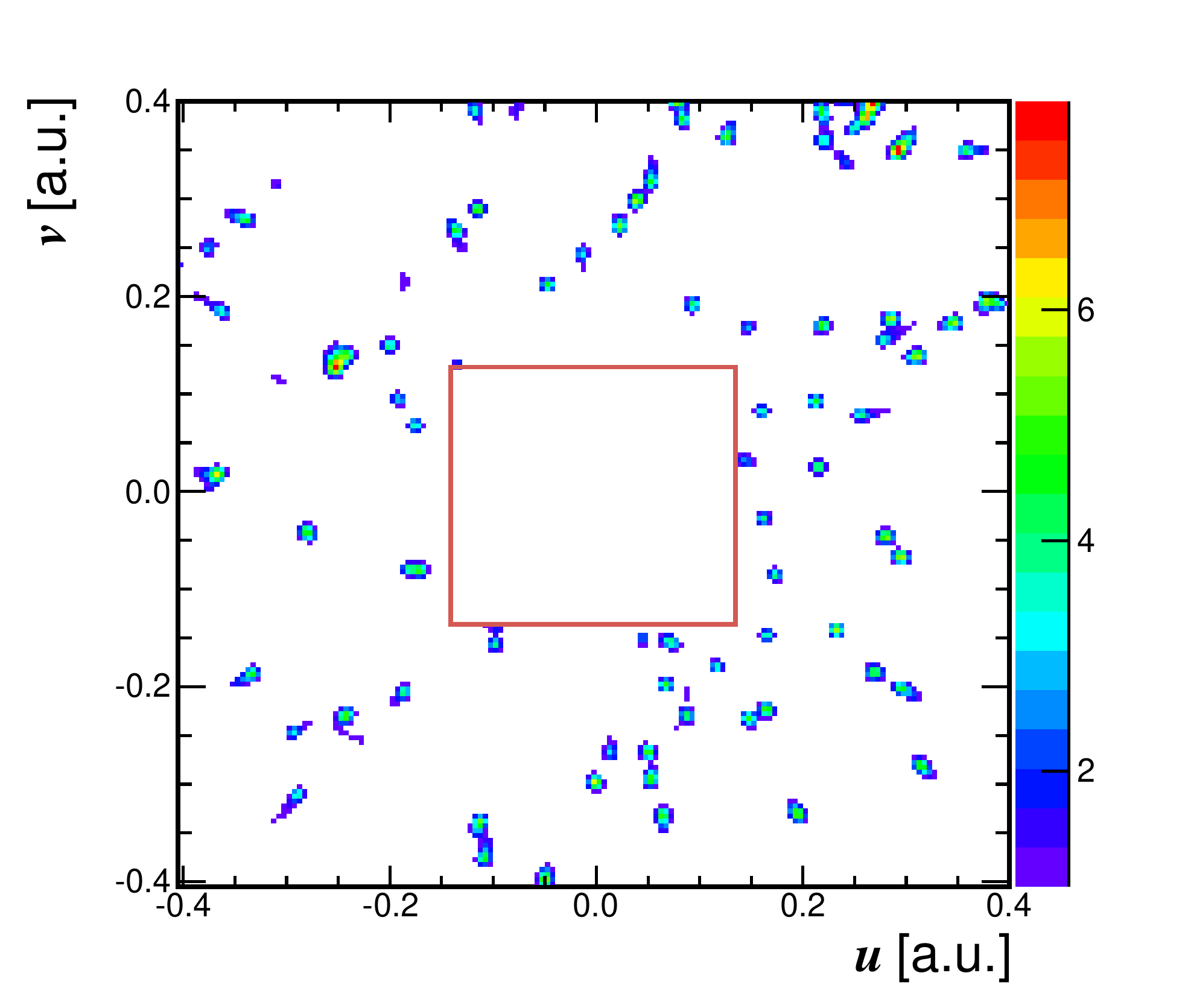} 
\includegraphics[width=.42\textwidth]{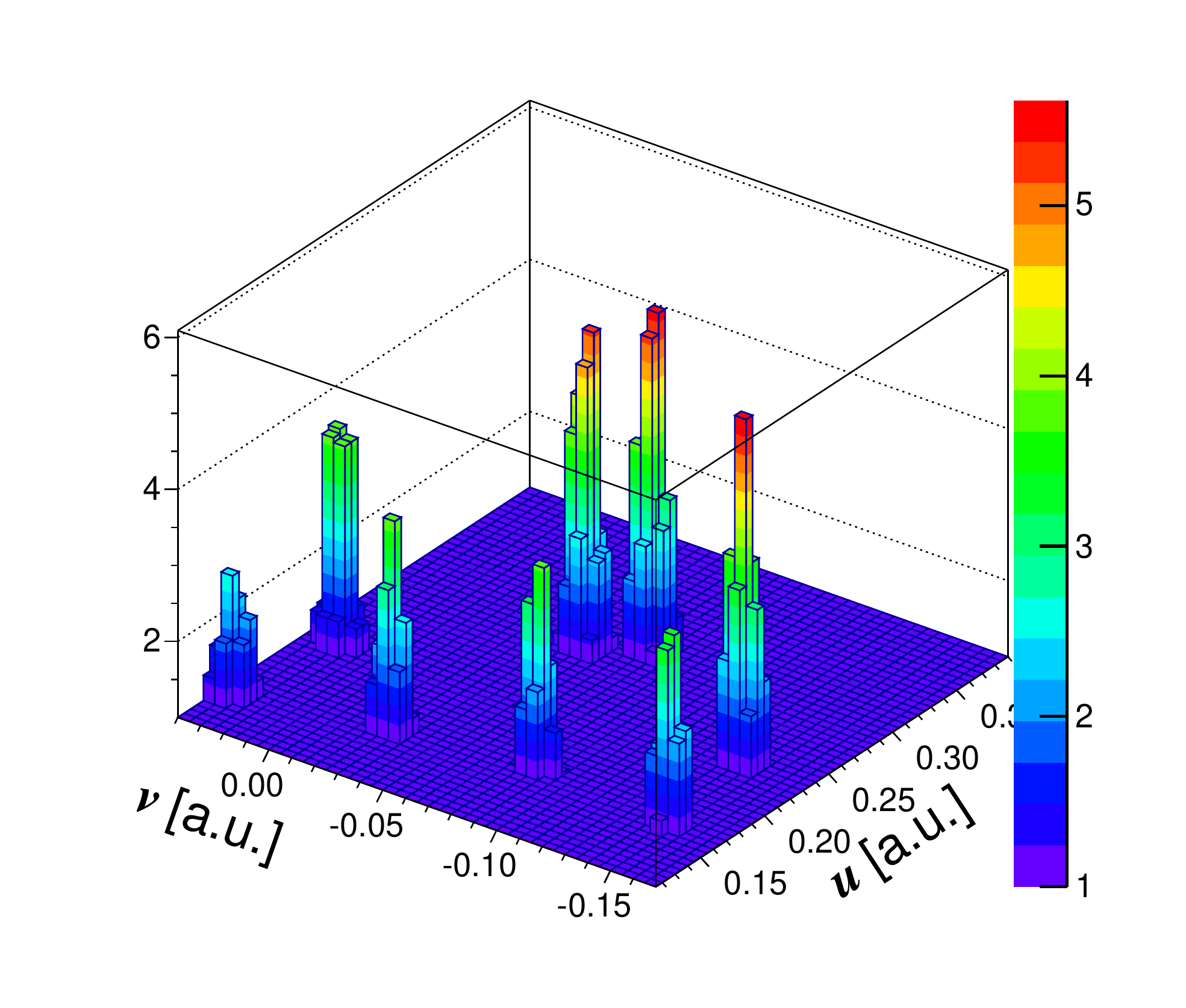} 
\caption{Left: response of the retina algorithm
(only the $(u,v)$-plane, where the pattern recognition is made) to
a minimum-bias event from the LHCb MC, 
with instantaneous luminosity of $L=2\times 10^{33} \, \rm cm^{-2} s^{-1}$. 
The hole at the centre of the figure is due to the physical hole in the VELO layers.
Right: a zoom of the retina response.}
\label{fig:ev_MC}
\end{figure}
Generic collision samples from the LHCb simulation are used to assess the performances
of the retina algorithm.
Events are generated with PYTHIA 8~\cite{PYTHIA}, 
with beam energy of $7\,\mathrm{TeV}$, in the two luminosity scenarios
expected for the LHCb-Upgrade operation:
(i) $L=2\times 10^{33} \, \rm cm^{-2} s^{-1}$, with an average number of interactions per 
bunch crossing equal to $7.6$; 
(ii) $L=3\times 10^{33} \, \rm cm^{-2} s^{-1}$, where the average number of interactions per 
bunch crossing is equal to $11.4$.
A typical response of the retina algorithm to a simulated event of the LHCb experiment
is shown in figure~\ref{fig:ev_MC}, where 
several clusters are clearly identifiable, 
and most of them reconstructed as tracks.

\begin{figure}[tbp]
\centering
\begin{overpic}[width=.45\textwidth]{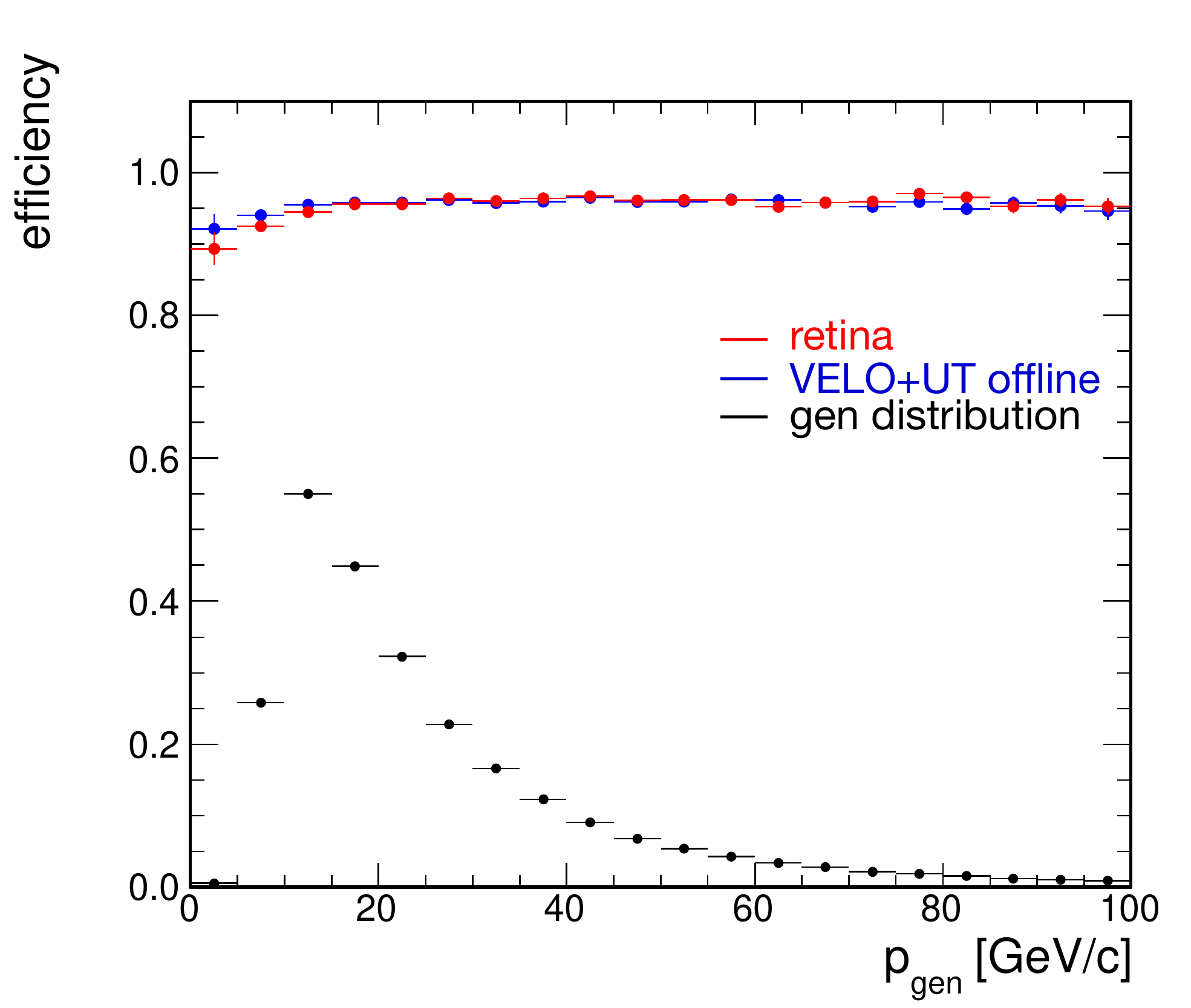}
	\put(85, 35){$(a)$}
\end{overpic}
\begin{overpic}[width=.45\textwidth]{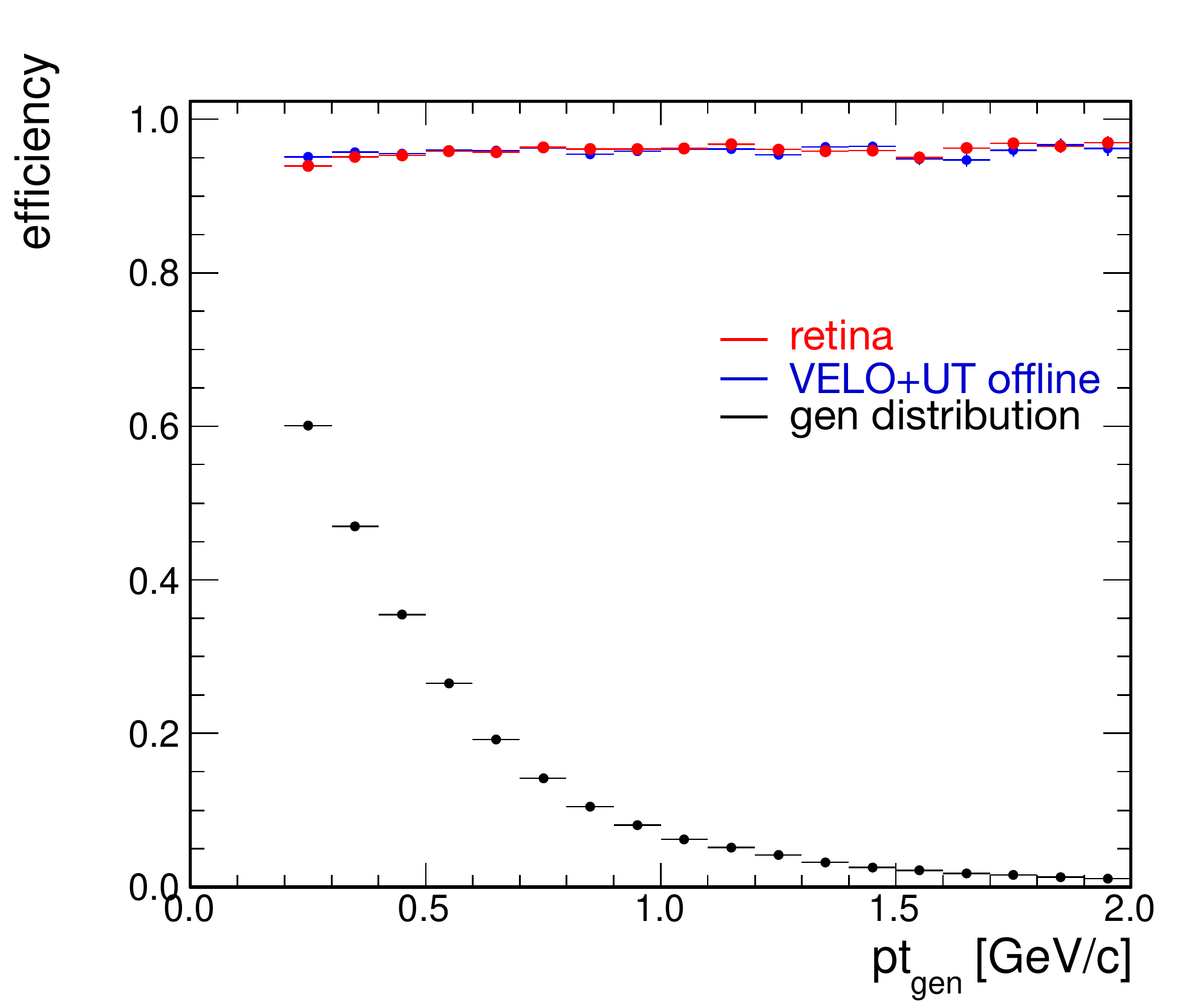}
	\put(85, 35){$(b)$}
\end{overpic} \\
\begin{overpic}[width=.45\textwidth]{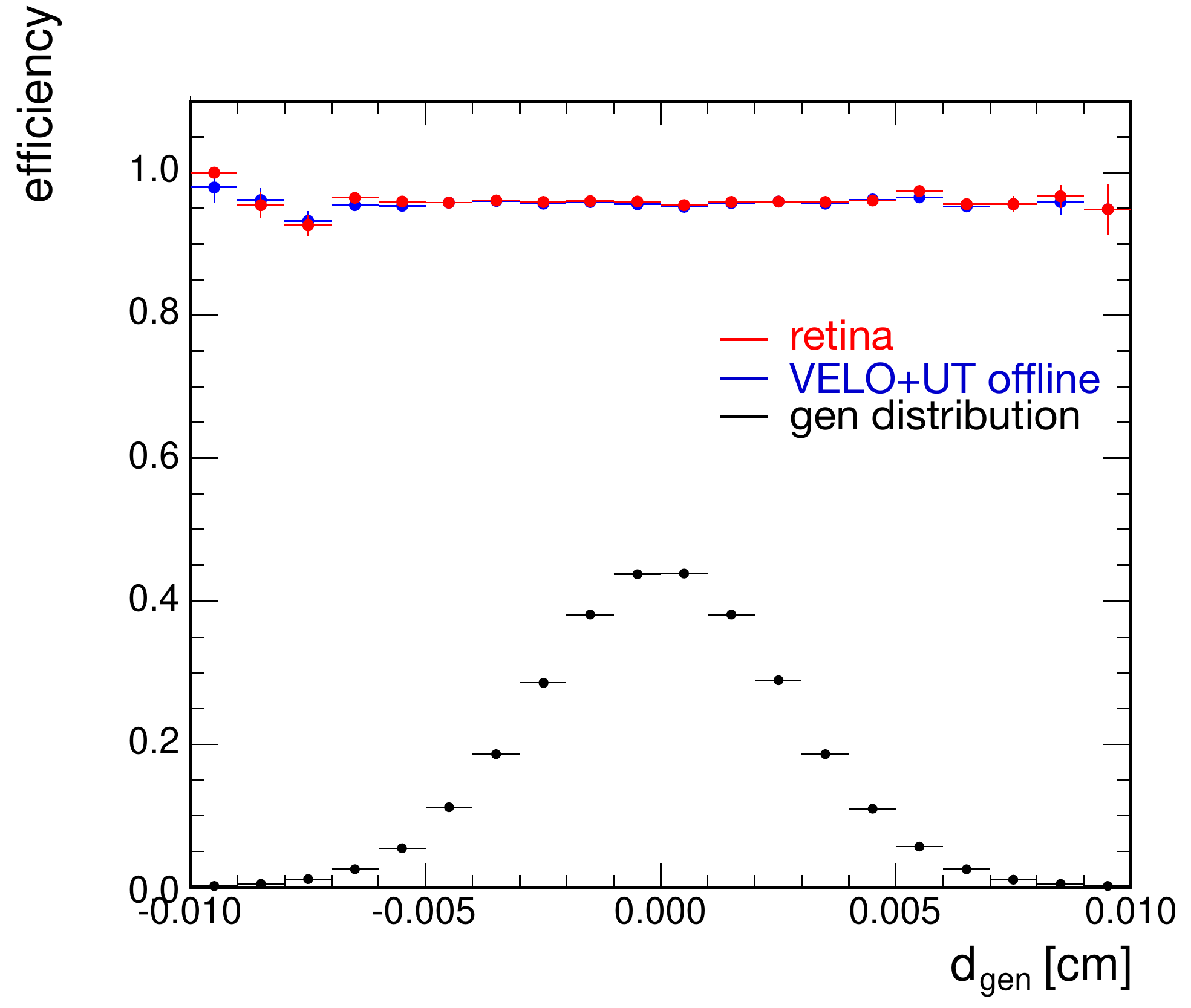}
	\put(85, 35){$(c)$}
\end{overpic} 
\begin{overpic}[width=.45\textwidth]{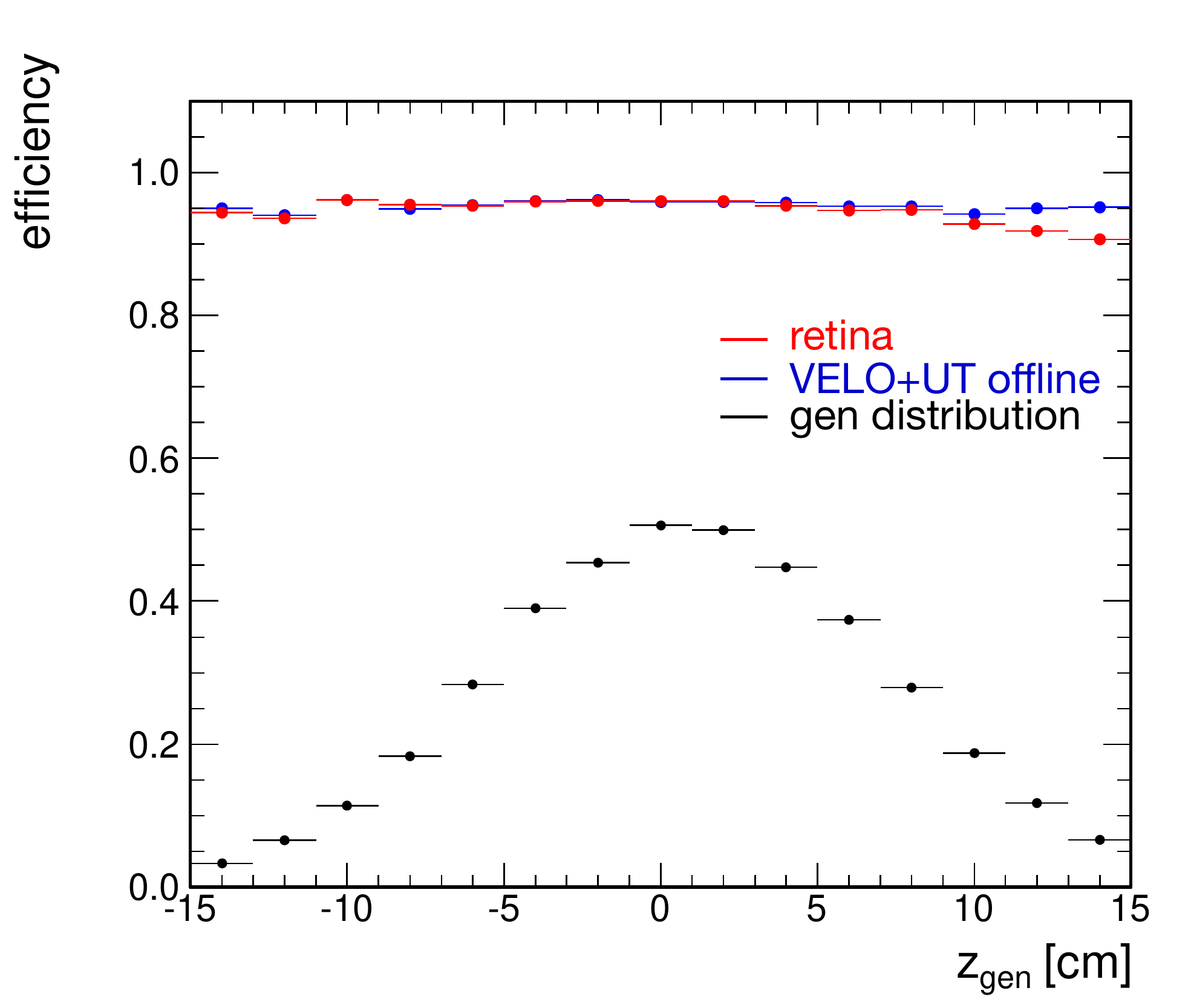}
	\put(85, 35){$(d)$}
\end{overpic}  
\caption{Tracking reconstruction efficiency of the retina algorithm (in red) and 
of the offline VELO+UT algorithm (in blue), as function
of: $(a)$ $p$, $(b)$ $p_T$, $(c)$ $d$, $(d)$ $z_0$. 
The distribution of the considered 
parameter is, also, reported in black.
Luminosity of $L=3\times 10^{33} \, \rm cm^{-2} s^{-1}$.}
\label{fig:efficiency_retina}
\end{figure}

To benchmark the retina performances we compare the retina algorithm with the 
the first stage of the LHCb-Upgrade tracking sequence, named VELOUT algorithm~\cite{VELOUT}. 
Since the chosen layer
configuration has a different acceptance with respect to the the VELOUT algorithm, 
we considered
only tracks in the region of the $(u,v)$-plane
corresponding approximatively to $\theta < 50 \,\rm mrad$.
In addition, we required at least three hits on VELO layers
and two hits on UT layers.
Further cuts on momentum $(p>3\,\mathrm{GeV}/c)$ 
and on transverse momentum  $(p_T>200\,\mathrm{MeV}/c)$ of the track are also applied.
Tracks satisfying these requirements are defined as
\emph{reconstructable},  and the 
tracking efficiency is defined as the number of reconstructed tracks 
over the number of reconstructable tracks. 
The efficiency of the retina is reported in figure~\ref{fig:efficiency_retina} 
as function of $p,\, p_T, \,d, \,z_0$ parameters. 
By comparison 
we also report  the efficiency  of VELOUT algorithm,
performing the same task as the retina~\cite{VELOUT, notaPubLHCb}.
%
%
The retina algorithm shows very
high efficiencies in reconstructing tracks,  95\% for minimum-bias tracks, which is comparable to the
offline tracking algorithm.  The fake track rate is
8\% at $L=2\times 10^{33} \, \rm cm^{-2} s^{-1}$ and 12\% at $L=3\times 10^{33} \, \rm cm^{-2} s^{-1}$,
at the same level of that obtained by the offline algorithm~\cite{VELOUT}. 
We also estimate the efficiency of the retina algorithm in
reconstructing signal tracks from some  benchmark decay modes, 
such as $B_s^0 \to \phi\phi$, $D^{*\pm} \to D^0\pi^{\pm}$ and $B^0 \rightarrow K^{*} \mu \mu$
for $L = 2\times 10^{33}\,\rm cm^{-2}s^{-1}$, which are reported in
table~\ref{tab:channels}. Similar results have been found at $3\times 10^{33}\,\rm cm^{-2}s^{-1}$.
\begin{table}[htb]
\centering
\caption{Retina efficiency on several benchmark channels.}
\label{tab:channels}
\begin{tabular}{l c }
\toprule
  & $2\times 10^{33}\rm cm^{-2}s^{-1}$  \\
\cmidrule{2-2}
		$B^0_s \rightarrow \phi \phi$ (signal tracks)		& 0.97 \\
		$D^{*+} \rightarrow D^{0} \pi^{+}$ (signal tracks)	& 0.97 \\
		$B^0 \rightarrow K^{*} \mu \mu$ (signal tracks)	 	& 0.98 \\
\bottomrule
\end{tabular}
\end{table}

\begin{figure}[tbp]
\centering
\includegraphics[width=.9\textwidth]{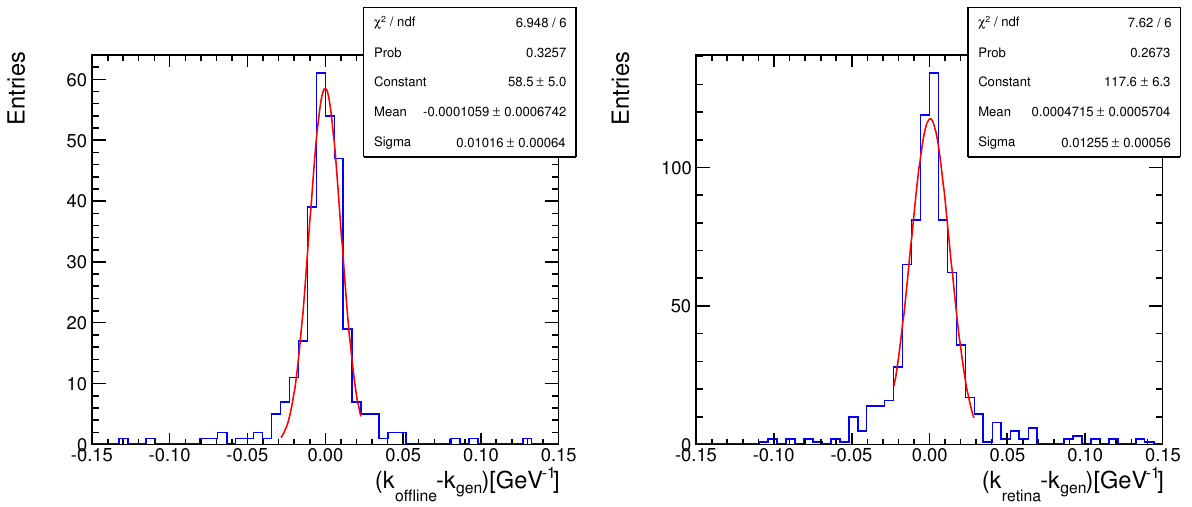}
\caption{Comparison between curvature resolution achieved with the retina (right panel) and with the VELOUT offline reconstruction algorithm (left panel).}
\label{fig:curvature_resolution}
\end{figure}

We also investigated track parameter resolution returned by the retina
algorithm. We found a resolution comparable to the offline algorithm,
taking into account the differences between our layers configuration
and  that one used by the official offline algorithm, which uses all
VELO layers and both axial and stereo UT layers. For instance the measurement of the track
curvature performed using the retina algorithm is less precise by a
factor $0.25$ (see figure~\ref{fig:curvature_resolution}), which is in agreement with the expected degradation due to the lack of the stereo layers
information in the retina algorithm~\cite{notaPubLHCb}. 
Among all track parameters, we focused our studies on the 
 curvature resolution, since it is the most important track parameter
 providing discriminating power between heavy flavor signal events and
 background events. Experience from past experiments, as CDF, and LHCb
 itself demonstrated that requirements on lifetime related
 quantities, as the impact parameter, are effective only if a
 requirement  on track momentum has been previously applied.  
 Preliminary studies have shown that no degradation in
 resolution is expected for the measurement of track parameters if the
 same number of detector layers is used for both algorithm.

\section{Conclusions}
We presented the first implementation of the artificial retina algorithm into a real
HEP environment like LHCb-Upgrade experiment. Performances of the algorithm 
were investigated.
Excellent performances have been found for the retina pattern-recognition.
Furthermore, it comes out that the retina has
a high efficiency on heavy quarks decays channels.
The retina algorithm is a powerful algorithm that exploits massive parallelism and
in conjunction with high-end FPGAs can provide a high-quality 
track reconstruction at the full LHC crossing rate~\cite{talk_Diego, notaPubLHCb}.


\end{document}